\documentclass{prop2015v2}
\usepackage[english]{babel}
\usepackage{epsfig, amsmath, amssymb,yfonts, fancyhdr}
\usepackage{amsthm}
\usepackage{mathrsfs}
\usepackage{graphicx}
\usepackage{mdwlist}
\usepackage{multirow}
\usepackage{marginnote}
\usepackage{booktabs}
\usepackage{titlesec}
\usepackage{appendix}
\usepackage{psfrag}
\usepackage{color}
\usepackage{tensor}
\usepackage{slashed}
\usepackage{setspace}
\usepackage{subfigure}

\usepackage{tikz}
\usetikzlibrary{matrix}
\usetikzlibrary{arrows}
\usetikzlibrary{fit}
\usetikzlibrary{backgrounds}
\usetikzlibrary{calc}
\usetikzlibrary{positioning}
\usetikzlibrary{scopes}
\usetikzlibrary{decorations.pathreplacing}
\usetikzlibrary{decorations.pathmorphing}
\usetikzlibrary{decorations.markings}
\usetikzlibrary{shapes}
\usetikzlibrary{fadings}
\usetikzlibrary{intersections}

\makeatletter
\begingroup
\def\IfPGFversionIIorIII{%
  \expandafter\getPGFv@rsion\pgfversion\@nil
}
\def\getPGFv@rsion#1.#2\@nil{%
  \ifnum#1<3
    \expandafter\endgroup\expandafter\@firstoftwo
  \else
    \expandafter\endgroup\expandafter\@secondoftwo
  \fi}
\IfPGFversionIIorIII
  {\pgfmathdeclarefunction{Xatan2}{2}{\csname pgfmathatan2\endcsname{#2}{#1}}}
  {\pgfmathdeclarefunction{Xatan2}{2}{\csname pgfmathatan2\endcsname{#1}{#2}}}
\makeatother

\makeatletter
\newdimen\@cx
\newdimen\@cy
\newdimen\@ax
\newdimen\@ay
\newdimen\@bx
\newdimen\@by

\newcommand{\findArcThroughPoints}[3]{
  \pgfextractx{\@cx}{\pgfpointanchor{#1}{center}}
  \pgfextracty{\@cy}{\pgfpointanchor{#1}{center}}
  \pgfextractx{\@ax}{\pgfpointanchor{#2}{center}}
  \pgfextracty{\@ay}{\pgfpointanchor{#2}{center}}
  \pgfextractx{\@bx}{\pgfpointanchor{#3}{center}}
  \pgfextracty{\@by}{\pgfpointanchor{#3}{center}}

  \pgfmathsetmacro{\radius}{veclen((\@cx) - (\@ax), (\@cy) - (\@ay))*0.0352777778}
  \pgfmathsetmacro{\startAngle}{Xatan2((\@ay)-(\@cy), (\@ax) - (\@cx))}
  \pgfmathsetmacro{\endAngle}{Xatan2((\@by)-(\@cy), (\@bx) - (\@cx))}
}
\makeatother

\tikzstyle{reverseclip}=[insert path={(current page.north east) --
  (current page.south east) --
  (current page.south west) --
  (current page.north west) --
  (current page.north east)}
]

\tikzset{
    set arrow inside/.code={\pgfqkeys{/tikz/arrow inside}{#1}},
    set arrow inside={end/.initial=>, opt/.initial=},
    /pgf/decoration/Mark/.style={
        mark/.expanded=at position #1 with
        {
            \noexpand\arrow[\pgfkeysvalueof{/tikz/arrow inside/opt}]{\pgfkeysvalueof{/tikz/arrow inside/end}}
        }
    },
    arrow inside/.style 2 args={
        set arrow inside={#1},
        postaction={
            decorate,decoration={
                markings,Mark/.list={#2}
            }
        }
    },
}

\tikzset{
        with arrow/.style={arrow inside={end=arrowhead}{#1}},
        with arrow/.default={0.5}
}

\newcommand\shiftedarrowhead[1]{
  \pgfmathsetlengthmacro{\originx}{#1\pgflinewidth + 2\pgflinewidth}
  \pgfpathmoveto{\pgfpoint{\originx}{0pt}}
  \pgfpathcurveto
  {\pgfpoint{\originx-3.3333\pgflinewidth}{.5\pgflinewidth}}
  {\pgfpoint{\originx-7\pgflinewidth}{2\pgflinewidth}}
  {\pgfpoint{\originx-10\pgflinewidth}{3.75\pgflinewidth}}
  \pgfpathlineto{\pgfpoint{\originx-10\pgflinewidth}{-3.75\pgflinewidth}}
  \pgfpathcurveto
  {\pgfpoint{\originx-7\pgflinewidth}{-2\pgflinewidth}}
  {\pgfpoint{\originx-3.3333\pgflinewidth}{-.5\pgflinewidth}}
  {\pgfpoint{\originx}{0pt}}
  \pgfusepathqfill
}

\newcommand\shiftedsplitarrowhead[1]{
\pgfmathsetlengthmacro{\originx}{#1\pgflinewidth + 2 \pgflinewidth}
  \pgfpathmoveto{\pgfpoint{\originx}{1\pgflinewidth}}
  \pgfpathcurveto
  {\pgfpoint{\originx-3.3333\pgflinewidth}{1.4\pgflinewidth}}
  {\pgfpoint{\originx-7\pgflinewidth}{2.6\pgflinewidth}}
  {\pgfpoint{\originx-10\pgflinewidth}{4\pgflinewidth}}
  \pgfpathlineto{\pgfpoint{\originx-10\pgflinewidth}{1\pgflinewidth}}
  \pgfpathlineto{\pgfpoint{\originx}{1\pgflinewidth}}
  \pgfusepathqfill
  \pgfpathmoveto{\pgfpoint{\originx}{-1\pgflinewidth}}
   \pgfpathcurveto
  {\pgfpoint{\originx-3.3333\pgflinewidth}{-1.4\pgflinewidth}}
  {\pgfpoint{\originx-7\pgflinewidth}{-2.6\pgflinewidth}}
  {\pgfpoint{\originx-10\pgflinewidth}{-4\pgflinewidth}}
  \pgfpathlineto{\pgfpoint{\originx-10\pgflinewidth}{-1\pgflinewidth}}
  \pgfpathlineto{\pgfpoint{\originx}{-1\pgflinewidth}}  
  \pgfusepathqfill
}

\pgfarrowsdeclare{arrowhead}{arrowhead}
{
  \pgfarrowsleftextend{0\pgflinewidth}
  \pgfarrowsrightextend{0\pgflinewidth}
}
{
  \shiftedarrowhead{4.5}
}

\pgfarrowsdeclare{split arrowhead}{split arrowhead}
{
  \pgfarrowsleftextend{0\pgflinewidth}
  \pgfarrowsrightextend{0\pgflinewidth}
}
{
  \shiftedsplitarrowhead{4.5}
}

\pgfarrowsdeclare{double arrowhead}{double arrowhead}
{
  \pgfarrowsleftextend{0\pgflinewidth}
  \pgfarrowsrightextend{0\pgflinewidth}
}
{
  \shiftedarrowhead{9}
  \shiftedarrowhead{0}
}

\pgfarrowsdeclare{split double arrowhead}{split double arrowhead}
{
  \pgfarrowsleftextend{0\pgflinewidth}
  \pgfarrowsrightextend{0\pgflinewidth}
}
{
  \shiftedsplitarrowhead{9}
  \shiftedsplitarrowhead{0}
}

\pgfarrowsdeclare{triple arrowhead}{triple arrowhead}
{
  \pgfarrowsleftextend{0\pgflinewidth}
  \pgfarrowsrightextend{0\pgflinewidth}
}
{
  \shiftedarrowhead{13.5}
  \shiftedarrowhead{4.5}
  \shiftedarrowhead{-4.5}
}

\pgfarrowsdeclare{split triple arrowhead}{split triple arrowhead}
{
  \pgfarrowsleftextend{0\pgflinewidth}
  \pgfarrowsrightextend{0\pgflinewidth}
}
{
  \shiftedsplitarrowhead{13.5}
  \shiftedsplitarrowhead{4.5}
  \shiftedsplitarrowhead{-4.5}
}

\pgfarrowsdeclare{quadruple arrowhead}{quadruple arrowhead}
{
  \pgfarrowsleftextend{0\pgflinewidth}
  \pgfarrowsrightextend{0\pgflinewidth}
}
{
  \shiftedarrowhead{16}
  \shiftedarrowhead{8}
  \shiftedarrowhead{0}
  \shiftedarrowhead{-8}
}

\pgfarrowsdeclare{split quadruple arrowhead}{split quadruple arrowhead}
{
  \pgfarrowsleftextend{0\pgflinewidth}
  \pgfarrowsrightextend{0\pgflinewidth}
}
{
  \shiftedsplitarrowhead{16}
  \shiftedsplitarrowhead{8}
  \shiftedsplitarrowhead{0}
  \shiftedsplitarrowhead{-8}
}

\pgfarrowsdeclare{quintuple arrowhead}{quintuple arrowhead}
{
  \pgfarrowsleftextend{0\pgflinewidth}
  \pgfarrowsrightextend{0\pgflinewidth}
}
{
  \shiftedarrowhead{20}
  \shiftedarrowhead{12}
  \shiftedarrowhead{4}
  \shiftedarrowhead{-4}
  \shiftedarrowhead{-12}
}

\pgfarrowsdeclare{split quintuple arrowhead}{split quintuple arrowhead}
{
  \pgfarrowsleftextend{0\pgflinewidth}
  \pgfarrowsrightextend{0\pgflinewidth}
}
{
  \shiftedsplitarrowhead{20}
  \shiftedsplitarrowhead{12}
  \shiftedsplitarrowhead{4}
  \shiftedsplitarrowhead{-4}
  \shiftedsplitarrowhead{-12}
}

\tikzset{
  on each segment/.style={
    decorate,
    decoration={
      show path construction,
      moveto code={},
      lineto code={
        \path [#1]
        (\tikzinputsegmentfirst) -- (\tikzinputsegmentlast);
      },
      curveto code={
        \path [#1] (\tikzinputsegmentfirst)
        .. controls
        (\tikzinputsegmentsupporta) and (\tikzinputsegmentsupportb)
        ..
        (\tikzinputsegmentlast);
      },
      closepath code={
        \path [#1]
        (\tikzinputsegmentfirst) -- (\tikzinputsegmentlast);
      },
    },
  },
  mid arrow/.style={#1, postaction={decorate,decoration={
        markings,
        mark=at position .5 with {\arrow[#1]{arrowhead}}
      }}},
  custom arrow style/.style 2 args={#2, postaction={decorate,decoration={
        markings,
        mark=at position #1 with {\arrow[#2]{arrowhead}}
      }}},
  double mid arrow/.style={#1, postaction={decorate,decoration={
        markings,
        mark=at position .5 with {\arrow[#1]{double arrowhead}}
      }}},
  triple mid arrow/.style={#1, postaction={decorate,decoration={
        markings,
        mark=at position .5 with {\arrow[#1]{triple arrowhead}}
      }}},
  quadruple mid arrow/.style={#1, postaction={decorate,decoration={
        markings,
        mark=at position .5 with {\arrow[#1]{quadruple arrowhead}}
      }}},
  quintuple mid arrow/.style={#1, postaction={decorate,decoration={
        markings,
        mark=at position .5 with {\arrow[#1]{quintuple arrowhead}}
      }}},
  double fundamental arrow/.style={#1, postaction={decorate,decoration={
        markings,
        mark=at position .3 with {\arrow[#1]{arrowhead}},
        mark=at position .7 with {\arrow[#1,rotate=180]{arrowhead}}
      }}},
  double antifundamental arrow/.style={#1, postaction={decorate,decoration={
        markings,
        mark=at position .3 with {\arrow[#1,rotate=180]{arrowhead}},
        mark=at position .7 with {\arrow[#1]{arrowhead}}
      }}},
  double antifundamental double arrow/.style={#1, postaction={decorate,decoration={
        markings,
        mark=at position .15 with {\arrow[#1,rotate=180]{arrowhead}},
        mark=at position .3 with {\arrow[#1,rotate=180]{arrowhead}},
        mark=at position .7 with {\arrow[#1]{arrowhead}},
        mark=at position .85 with {\arrow[#1]{arrowhead}}
      }}},
}

\tikzset{single arrow/.style={on each segment={mid arrow={#1}}}}
\tikzset{single arrow/.default={quiver path}}
\tikzset{custom arrow/.style 2 args={on each segment={custom arrow style={#1}{#2}}}}
\tikzset{double arrow/.style={on each segment={double mid arrow={#1}}}}
\tikzset{double arrow/.default={quiver path}}
\tikzset{triple arrow/.style={on each segment={triple mid arrow={#1}}}}
\tikzset{triple arrow/.default={quiver path}}
\tikzset{quadruple arrow/.style={on each segment={quadruple mid arrow={#1}}}}
\tikzset{quadruple arrow/.default={quiver path}}
\tikzset{quintuple arrow/.style={on each segment={quintuple mid arrow={#1}}}}
\tikzset{quintuple arrow/.default={quiver path}}
\tikzset{double fundamental/.style={on each segment={double fundamental
    arrow={#1}}}}
\tikzset{double fundamental/.default={quiver path}}
\tikzset{double antifundamental/.style={on each segment={double antifundamental
    arrow={#1}}}}
\tikzset{double antifundamental/.default={quiver path}}
\tikzset{double antifundamental double/.style={on each segment={double antifundamental
    double arrow={#1}}}}
\tikzset{double antifundamental double/.default={quiver path}}

\tikzset{double line/.style={on each segment={#1, double distance=1pt}}}
\tikzset{double line/.default={quiver path}}

\tikzset{double line arrow/.style={on each segment={#1, double distance=1pt, postaction={decorate,decoration={
        markings,
        mark=at position .5 with {\arrow[#1, scale=0.4]{split arrowhead}}
      }}}}}
\tikzset{double line arrow/.default={quiver path}}

\tikzset{double line double arrow/.style={on each segment={#1, double distance = 1pt, postaction={decorate,decoration={
        markings,
        mark=at position .5 with {\arrow[#1, scale=0.4]{split double arrowhead}}
      }}}}}
\tikzset{double line double arrow/.default={quiver path}}

\tikzset{double line triple arrow/.style={on each segment={#1, double distance = 1pt, postaction={decorate,decoration={
        markings,
        mark=at position .5 with {\arrow[#1, scale=0.4]{split triple arrowhead}}
      }}}}}
\tikzset{double line triple arrow/.default={quiver path}}

\tikzset{double line quadruple arrow/.style={on each segment={#1, double distance = 1pt, postaction={decorate,decoration={
        markings,
        mark=at position .5 with {\arrow[#1, scale=0.4]{split quadruple arrowhead}}
      }}}}}
\tikzset{double line quadruple arrow/.default={quiver path}}

\tikzset{SU/.style={circle,quiver path, fill=yellow!20, minimum
    size=0.7cm}}
\tikzset{Sp/.style={circle split,rotate=90, quiver path, fill=yellow!20, minimum
    size=0.7cm}}
\tikzset{SO/.style={circle split,quiver path, fill=yellow!20, minimum
    size=0.7cm}}
\tikzset{Flavor/.style={rectangle, quiver path, fill=blue!20, minimum
    size=0.55cm}}
\tikzset{asymm/.style={rectangle split, rectangle split parts=2,
    rectangle split empty part width=0.1, fill=white,
    xscale=0.5, yscale=0.8,
    draw=black, minimum size=0.7cm}}
\tikzset{symm/.style={asymm, rectangle split horizontal,
    xscale=0.8/0.5, yscale=0.5/0.8}}

\tikzset{quiver path/.style={draw=black}}

\tikzset{quiver/.style={auto, node distance=2em, thick, x=5em, y=5em,
                              every path/.style=quiver path}}

\tikzset{toric node/.style={draw,circle,inner sep=1.5pt,fill}}
\tikzset{toric line/.style={draw, thick}}
\tikzset{toric diagram/.style={scale=0.7,
               every node/.style=toric node,
               every path/.style=toric line}}

\tikzset{superpotential node/.style={circle, draw=black}}
\tikzset{white/.style={superpotential node, fill=white}}
\tikzset{black/.style={superpotential node, fill=black}}
\tikzset{no unit cell/.style={background rectangle/.style={draw=none}}}
\tikzset{unit cell/.style={draw=gray, dashed}}
\tikzset{dimer/.style={auto, draw, thick, framed, tight background,
                       background rectangle/.style=unit cell,
                       every node/.style={inner sep=2pt},
                       every path/.style={}}}

\tikzset{5-brane diagram/.style={scale=1.7,auto, draw, thick, tight background,
                                 every path/.style={draw, thick}}}
\tikzset{cw/.style={draw=none, fill=orange!20}}
\tikzset{acw/.style={draw=none, fill=gray!30}}
\tikzset{O5/.style={star, star point ratio=2, fill=yellow, draw=black, scale=0.5}}

\tikzset{O5charges/.style={matrix of math nodes, every node/.style={scale=0.8}}}

\tikzset{tiling intersection/.style={draw, fill=yellow!20, thin, circle, scale=0.6}}
\tikzset{flavor intersection/.style={draw=black, fill=blue!20,
thin, regular polygon, regular polygon sides=4, scale=0.6}}

\tikzset{phantom edge/.style={draw, opacity=0.3}}

\title{On the physical realization of Seiberg duality for branes at the $\mathbb{F}_0$ singularity}
\author[M. Berasaluce-Gonz\'alez]{Mikel Berasaluce-Gonz\'alez\inst{1,}\footnote{E-mail:~\textsf{mberasal@uni-mainz.de}}}
\address[1]{Institute for Physics (WA THEP) $\&$ Cluster of Excellence PRISMA, Johannes Gutenberg University, D-55099 Mainz}
\begin{abstract}
Branes at a $\mathbb{F}_0$ singularity give rise to two different toric quiver gauge theories, which are related by Seiberg duality. We study where in the K\"ahler moduli space each of them is physically realized. 
\end{abstract}
\shortabstract
\begin{document}
\maketitle

\section{Introduction}

Seiberg duality \cite{Seiberg:1994pq} is an important aspect of $\mathcal{N}=1$ supersymmetric gauge theories. It relates two theories which are different in the UV but flow to the same IR fixed point under a renormalization group flow. In the context of string theory, it is often realized by supersymmetric deformations of systems of branes that induce irrelevant deformations of the low energy effective field theory.

In the context of chiral quiver theories, the algebraic content of Seiberg duality can be understood at the level of topological string theory \cite{Berenstein:2002fi}. An interesting question is finding to what extent the same occurs in the full string theory, once the BPS conditions for the brane system are taken into account.

In this paper we will focus on the theories that arise un the worldvolume of branes located at a $\mathbb{F}_0$ singularity inside a Calabi-Yau threefold, since for this geometry there are two different toric quiver gauge theories that can arise, which are related by Seiberg duality. In this case, the BPS condition corresponds to all the periods characterizing the BPS phase of the fractional branes being aligned \cite{Aspinwall:2004vm}.

\section{The complex cone over $\mathbb{F}_0$}

The complex cone over $\mathbb{F}_0$ is a toric variety \cite{Cox:2000vi}. The toric diagram associated to its fan, its Mori cone and its Newton polynomial are shown in figure \ref{F0}.
\begin{figure}
  \begin{minipage}[c]{0.2\textwidth}
  \begin{subfigure}
  \centering
   \begin{tikzpicture}[toric diagram, scale=1]
    \node[label=above right:$t$] at (0,0) (t) {};
    \node[label=right:$x_2$] at (1,0) (x3) {};
    \node[label=above:$x_3$] at (0,1) (x4) {};
    \node[label=below:$x_1$] at (0,-1) (x2) {};
    \node[label=left:$x_4$] at (-1,0) (x5) {};

    \draw (x2) -- (x3) -- (x4) -- (x5);

    \path (t) -- (x3) (t) -- (x5) (t) -- (x4) (t) -- (x2) (x2) -- (x5);
\end{tikzpicture}
  \end{subfigure}
  \end{minipage}%
  \begin{minipage}[c]{0.5\textwidth}
     \begin{equation}\nonumber
      \begin{array}{c|cccccc}
      & x_1 & x_2 & x_3 & x_4 & t\\
      \hline
      \mathcal{C}_1 & 0 & 1 & 0 & 1 & -2\\
      \mathcal{C}_2 & 1 & 0 & 1 & 0 & -2
 \end{array}
 \end{equation}
\begin{equation}
  P(x,y)=\frac{a}{x}+\frac{b}{y}+cx+dy+t\nonumber
\end{equation}
\end{minipage}

\caption{Toric diagram, Mori cone and Newton polynomial for the complex cone over $\mathbb{F}_0$.}
\label{F0}
\end{figure}

The periods are computed by solving the Picard-Fuchs equations
\begin{subequations}
\begin{eqnarray}
 \theta_1^2-z_1(2\theta_1+2\theta_2)(2\theta_1+2\theta_2+1)\Phi(z_1,z_2)=0,\\
 \theta_2^2-z_2(2\theta_1+2\theta_2)(2\theta_1+2\theta_2+1)\Phi(z_1,z_2)=0,
 \end{eqnarray}
 \end{subequations}
where $\theta_i=z_i(\partial/\partial z_i)$, $z_1=bd/t^2$, $z_2=ac/t^2$.

After doing the analytic continuation of the periods, the quiver locus we obtain is\cite{Aspinwall:2004vm}
\begin{equation}
 z_1,z_2\rightarrow\infty,\quad\quad\frac{z_1}{z_2}=e^{i\alpha}.
\end{equation}

The mirror picture \cite{Cachazo:2001sg} is a double fibration over a complex plane W:
\begin{equation}
 W=uv,\quad W=P(x,y),\quad u,v\in\mathbb{C},\quad x,y\in\mathbb{C}^*.
\end{equation}
The first fiber is singular at the origin while the second one is singular at four points. The branes extend from the origin to each of these singular points while wrapping cycles in the fibers.

\subsection{Phases of the theory}

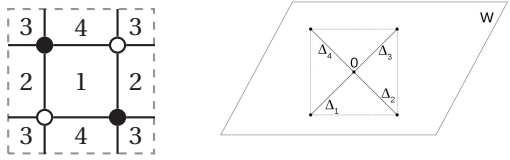
\begin{figure} 
\begin{subfigure}
  \centering
  \begin{tikzpicture}[dimer, no unit cell, x=1.5em, y=1.5em,
    remember picture]
    \def\drawdimer{
      \path
      (0,0) node[black] (1) {}
      ++(2,0) node[white] (2) {}
      ++(0,-2) node[black] (3) {}
      ++(-2,0) node[white] (4) {}
      ($(1) + (-2,0)$) node (l1) {}
      ($(4) + (-2,0)$) node (l2) {}
      ($(2) + (2,0)$) node (r1) {}
      ($(3) + (2,0)$) node (r2) {}
      ($(1) + (0,2)$) node (u1) {}
      ($(2) + (0,2)$) node (u2) {}
      ($(4) + (0,-2)$) node (d1) {}
      ($(3) + (0,-2)$) node (d2) {}
      
      ($(4) + (1,1)$) node {1}
      ($(3) + (0.5,1)$) node {2}
      ($(4) + (-0.5,1)$) node {2}
      ($(3) + (0.5,-0.5)$) node {3}
      ($(4) + (-0.5,-0.5)$) node {3}
      ($(1) + (-0.5,0.5)$) node {3}
      ($(2) + (0.5,0.5)$) node {3}
      ($(1) + (1,0.5)$) node {4}
      ($(4) + (1,-0.5)$) node {4}
      ;

      \draw
      (l1) -- (1) -- (2) -- (r1)
      (l2) -- (4) -- (3) -- (r2)
      (d1) -- (4) -- (1) -- (u1)
      (d2) -- (3) -- (2) -- (u2)
      ;
      }
    \begin{scope}
      \draw[unit cell]
        (-1,-3) -- (-1,1) -- (3,1) -- (3,-3) -- (-1, -3)
      ;
      \begin{pgfinterruptboundingbox}
      \clip
        (-1,-3) -- (-1,1) -- (3,1) -- (3,-3) -- (-1, -3)
      ;
      \end{pgfinterruptboundingbox}
      \drawdimer
    \end{scope}
  \end{tikzpicture}
\end{subfigure}
 \includegraphics[width=0.49\linewidth]{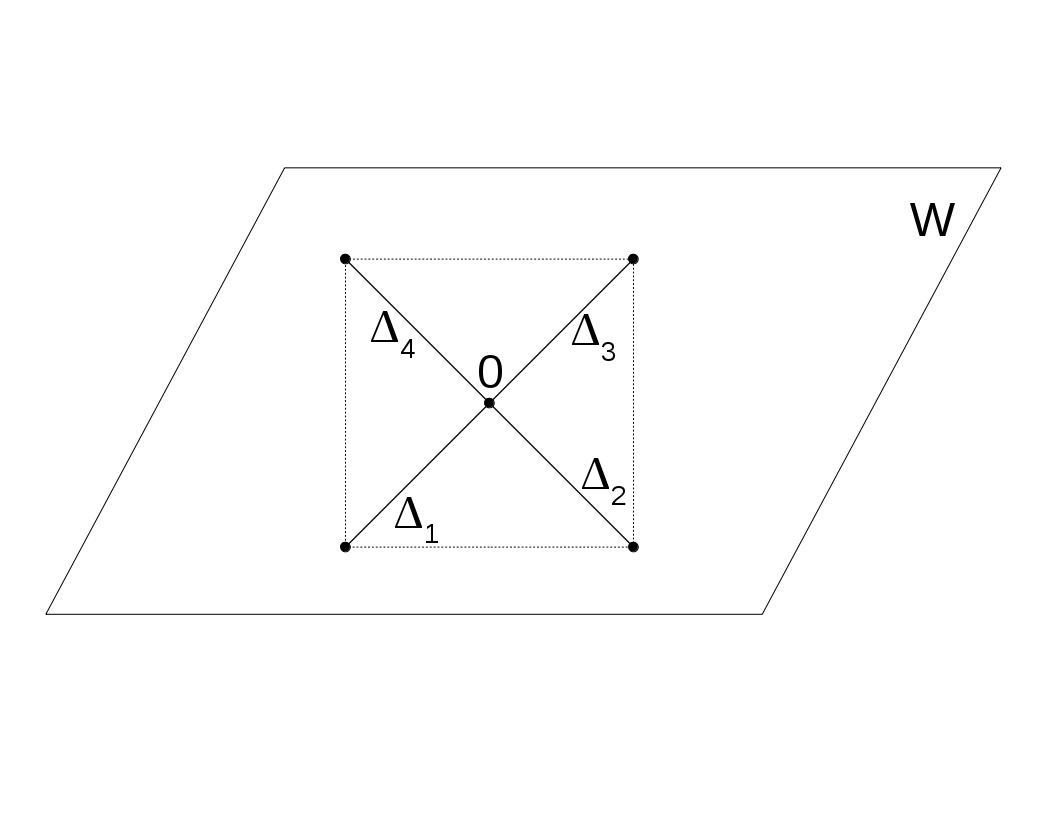}
\caption{Dimer and mirror picture for the phase I of $\mathbb{F}_0$.}
\label{F0_phase_I}
\end{figure}

\begin{figure} 
\begin{subfigure}
  \centering
  \begin{tikzpicture}[dimer, no unit cell, x=1em, y=1em,
    remember picture]
    \def\drawdimer{
      \path
      (0,0) node[black] (1) {}
      ++(2,0) node[white] (2) {}
      ++(0,2) node[black] (3) {}
      ++(-2,0) node[white] (4) {}
      ++(-1,1) node[black] (5) {}
      ++(4,0) node[white] (6) {}
      ++(0,-4) node[black] (7) {}
      ++(-4,0) node[white] (8) {}
      ($(5) + (-2,0)$) node (l1) {}
      ($(8) + (-2,0)$) node (l2) {}
      ($(6) + (2,0)$) node (r1) {}
      ($(7) + (2,0)$) node (r2) {}
      ($(5) + (0,2)$) node (u1) {}
      ($(6) + (0,2)$) node (u2) {}
      ($(8) + (0,-2)$) node (d1) {}
      ($(7) + (0,-2)$) node (d2) {}
      
      ($(1) + (1,1)$) node {1}
      ($(1) + (1,-1)$) node {4}
      ($(4) + (1,1)$) node {4}
      ($(2) + (1,1)$) node {2}
      ($(1) + (-1,1)$) node {2}
      ($(5) + (-0.5,0.5)$) node {3}
      ($(6) + (0.5,0.5)$) node {3}
      ($(7) + (0.5,-0.5)$) node {3}
      ($(8) + (-0.5,-0.5)$) node {3}
      ;

      \draw
      (1) -- (2) -- (3) -- (4) -- (1)
      (l1) -- (5) -- (u1)
      (r1) -- (6) -- (u2)
      (r2) -- (7) -- (d2)
      (l2) -- (8) -- (d1)
      (1) -- (8)
      (2) -- (7)
      (3) -- (6)
      (4) -- (5)
      ;
      }
    \begin{scope}
      \draw[unit cell]
        (-2,-2) -- (-2,4) -- (4,4) -- (4,-2) -- (-2, -2)
      ;
      \begin{pgfinterruptboundingbox}
      \clip
        (-2,-2) -- (-2,4) -- (4,4) -- (4,-2) -- (-2, -2)
      ;
      \end{pgfinterruptboundingbox}
      \drawdimer
    \end{scope}
  \end{tikzpicture}
\end{subfigure}
\includegraphics[width=0.5\linewidth]{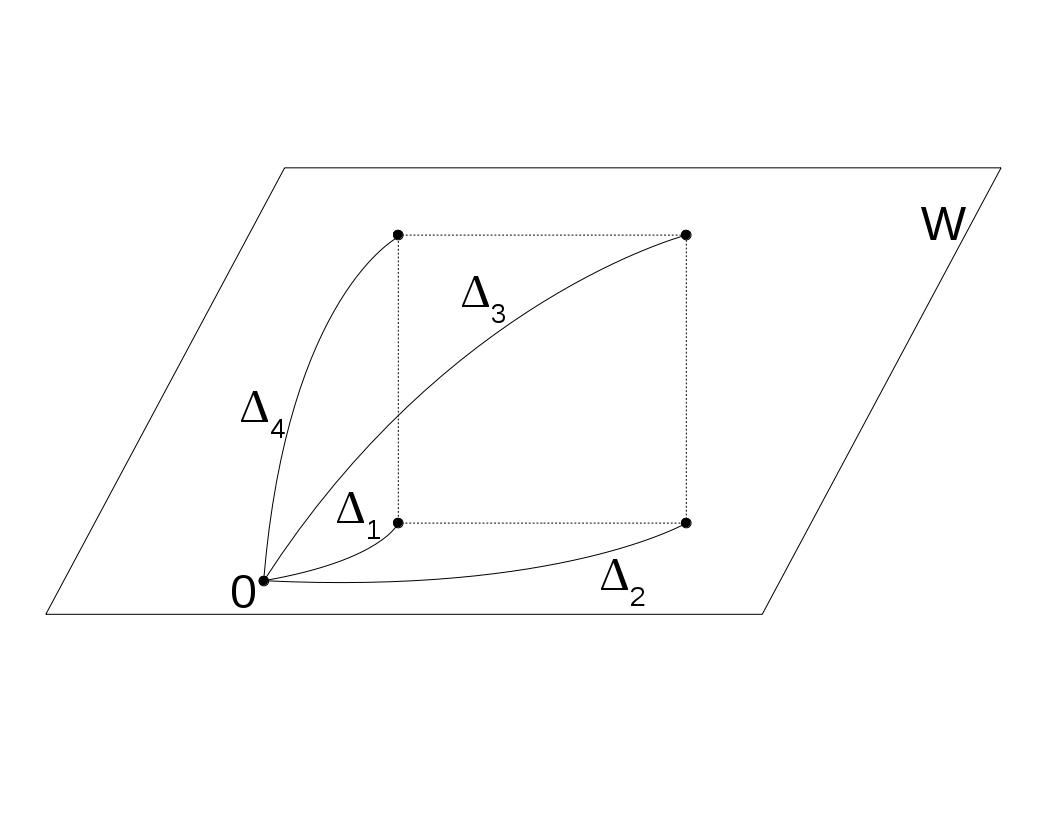}
\caption{Dimer and mirror picture for the phase II of $\mathbb{F}_0$.}
\label{F0_phase_II}
\end{figure}
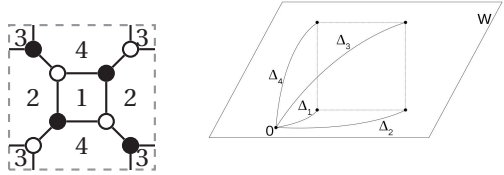

There are two different toric quiver gauge theories associated to the complex cone over $\mathbb{F}_0$, which we will simply denote as phase I and phase II. Figures \ref{F0_phase_I} and \ref{F0_phase_II} show the dimer and the mirror picture where, for simplicity, only the base with the points where the fibers become singular and the projection of the branes stretching between them is shown, for phase I and phase II, respectively. See \cite{GarciaEtxebarria:2006aq} for a review on dimers and how they encode the gauge group, field content and superpotential of a quiver gauge theory; and \cite{Cachazo:2001sg} for details of each of the phases and their mirror pictures.


\section{Physical realization of the different quiver gauge theories}
\subsection{Phase I}
\begin{figure}
 \begin{center}
  \includegraphics[width=0.6\linewidth]{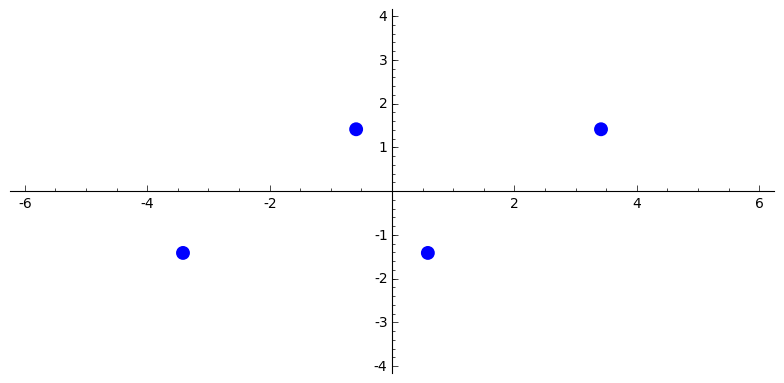}
  \caption{Mirror picture for the values of the paramenters indicated above.}
  \label{Phase_I_realization}
 \end{center}
\end{figure}

Phase I is physically realized at weak gauge coupling. For instance, taking $a=b=c=1$, $d=i$, $t\rightarrow 0$ (i.e., $z_1,z_2\rightarrow\infty$, $z_1/z_2=i$) we get the mirror picture shown in figure \ref{Phase_I_realization}, which corresponds with the one in figure \ref{F0_phase_I}.
\subsection{Phase II}
In the mirror picture, Seiberg duality corresponds to one of the points where the fiber is singular going through the origin of the base $W$ \cite{Cachazo:2001sg}. As we show in \cite{Berasaluce-Gonzalez:2015}, when $z_1=z_2$ while moving in the quiver locus two of the singular points coincide at the origin. Therefore, at the intersection of the conifold locus with the quiver locus there are precisely {\it two} branes becoming massless simultaneously. This implies that phase II will always be realized at strong gauge coupling, and never at weak gauge coupling.

\section{Conclusions}

We have studied the physical realization of the two toric gauge theories that may arise when we put D-branes in a singularity that locally is a complex cone over $\mathbb{F}_0$. The one we call phase I is realized at weak gauge coupling, but phase II is only realized in the strong coupling limit.

{\bf Acknowledgements:} It is a pleasure to thank I. Garc\'ia-Etxebarria and B. Heidenreich for collaboration on the work presented here. This work is partially supported by the DFG research grant HO 4166/2-1 and the {\it Cluster of Excellence PRISMA} DFG no. EXC 1098.

\end{document}